\documentclass{ws-ijmpa-1}
\usepackage[super,compress]{cite}
\usepackage{graphicx}

\usepackage{hyperref}
\usepackage{color}

\newcommand{\be}{\begin{equation}}
\newcommand{\ee}{\end{equation}}

\begin{document}
\markboth{N.V. Kolomoyets \& V.V. Skalozub}{Color Structure of Gluon Field Magnetic Mass}

%
\catchline{}{}{}{}{}
%

\title{Color Structure of Gluon Field Magnetic Mass}

\author{Natalia V. Kolomoyets\textsuperscript{*}}
\author{Vladimir V. Skalozub\textsuperscript{$\dag$}}

\address{Theoretical Physics Department, Oles Honchar Dnipro National University, Gagarin ave. 72\\
Dnipro, 49010, Ukraine\\
\textsuperscript{*}rknv7@mail.ru\\
\textsuperscript{$\dag$}skalozubvv@gmail.com}

\maketitle

\begin{history}
\phantom{\received{Day Month Year}}
\phantom{\revised{Day Month Year}}
\end{history}

\begin{abstract}
The color structure of the gluon field magnetic mass is investigated in the lattice SU(2) gluodynamics.  To realize that, the interaction between a monopole-antimonopole string and external neutral Abelian chromomagnetic field flux is considered. The string is introduced in the way proposed by Srednicki and Susskind. The neutral Abelian field flux is introduced through the twisted boundary conditions. Monte Carlo simulations are performed on 4D lattices at finite temperature. It is shown that the presence of the Abelian field flux weakens the screening of the string field. That means decreasing the gluon magnetic mass for this environment.  The contribution of the neutral Abelian field has the form of ``enhancing'' factor in the fitting functions.  This behavior independently confirms the long-range nature of the neutral Abelian field reported already in the literature.  The comparison
with analytic calculations is given. 
\end{abstract}.

\ccode{PACS numbers: 10.11.15.Ha}


\section{Introduction}

Investigation of new matter state -- quark-gluon plasma~(QGP),
which has to be  created at high temperature after the
deconfinement phase transition, is in the focus of modern high
energy physics. The critical (deconfinement) temperature $T_d$  of
the phase transition is estimated to be  160 --
180~MeV\cite{KS:lit:ABDFKKSK}. So it is expected that the plasma
will be discovered in experiments on collisions of protons or
heavy ions at Large Hadron Collider (LHC). Above this temperature,
the quarks and gluons are deliberated from hadrons and various
colored states (quantum or classical)   have to be present in QGP.
In particular,  the deliberation of the gluons leads to generation
of the gauge fields. As it is shown in Ref.~\citen{KS:lit:GPY},
these fields are screened. To describe screening, the magnetic,
$m_{magn}$ and electric~(Debye), $m_D,$ masses are introduced.
They are the main macroscopic characteristics of gauge fields at
finite temperature and  enter the correlators in the form $\sim
e^{- m_{magn} |\vec{r}|}, e^{- m_{magn} |\vec{r}|}, e^{- m_{D}
|\vec{r}|}$ which  show how fast the magnetic or electric
component of the field strengths decreases in space, $\vec{r}$ is
radius vector in space. In analytic calculations in imaginary
time formalism, these masses can be calculated from  gluon
polarization tensor $\Pi_{\mu\nu}(k_4,\vec{k})$ as the static
limit, $k_4 = 0, \vec{k} \to 0$ of the fourth components
$\Pi_{44}(k_4,\vec{k})$ (Debye mass), and $\Pi_{i i}(k_4,\vec{k})$
space components (magnetic mass). Here $k_\mu ( \mu, \nu = 1, 2,
3, 4)$ is Euclidean momentum of the field. As numerous studies
showed, the Debye mass squared is of the order $m_D^2 \sim g^2
T^2$ whereas the magnetic one is much smaller, $m_{magn}^2 \sim
g^4 T^2$, for small gauge coupling $g$; $T$ is temperature. Due to
the smallness of the latter  mass, it is not a simple problem to
estimate it precisely. Moreover, the color structure of it (its
analytic dependence on the presented fields) is not determined
finally, yet.

 In~Ref.~\citen{KS:lit:S} it was derived by analytic
methods of field theory  that magnetic mass of non-Abelian color
charged fields in the presence of constant neutral (Abelian)
chromomagnetic field $H$= const, described by the potential
$\bar{A}^a_\mu = H \delta^{a3} x^1 \delta_{\mu 2} $, where $a$ is
internal index, has the following structure:
\begin{equation}
    \label{KS:eq:mcolan}
    m^2_{col} \sim g^2T\sqrt{gH} ,
\end{equation}
where $m_{col}$ is the magnetic mass of the
color  charged chromomagnetic field. This field is defined as
$W_\mu^{\pm} = \frac{W^1_\mu \pm W^2_\mu}{\sqrt{2}}$, where 1, 2
are internal space indexes. The term ``Abelian constant
chromomagnetic field'' is also known as ``covariantly constant one''.
We will use the former term in what follows.

Another important point is the kind of magnetic fields considered.
In Ref.~\citen{KS:lit:ABDS} it was shown that this  Abelian
constant color magnetic field, being the solution of gauge field
equations without sources, has zero magnetic mass. This is like
usual static magnetic field in QED plasma. As it was realized
recently\cite{KS:lit:SB, KS:lit:SS, KS:lit:SM2017}, such kind of fields has to be
spontaneously generated in QGP and occupy the all plasma volume.
This field has to influence properties of
plasma~\cite{KS:lit:CC_SU2, KS:lit:CC_SU3}. For the spontaneously
generated field
\begin{equation}
    \sqrt{gH}\sim g^2T.
\end{equation}
Substitution of this field strength into Eq.\,(\ref{KS:eq:mcolan})
gives the known dependence of the magnetic mass on the coupling
constant and temperature.

In the present research we make a step in determination of
magnetic  mass structure for the SU(2) gluodynamics. Unlike the
Abelian chromomagnetic field, the non-Abelian one (related with
charged components $W^{\pm}_\mu$ which are self-interacting in the
chosen representation) cannot be  introduced on the lattice
directly through the twisted boundary conditions (TBC)
\cite{KS:lit:ABDS}.

One of the  ways  to detect  magnetic mass on the lattice  is to
introduce a monopole-antimonopole string  and investigate behavior
of its field. In Ref.~\citen{KS:lit:DGT} it was shown that the
field of the string is screened. However, in that paper the SU(2)
constituents were not distinguished.  The investigated field was
taken as the sum of both color neutral and  charged ones. So that
it is impossible to determine the contribution of each field
component to $ m_{magn}$.

In the present paper the methods applied in
Ref.~\citen{KS:lit:ABDS}  and Ref.~\citen{KS:lit:DGT} are joined
to separate  the comtributions of the  neutral Abelian and the charged
non-Abelian chromomagnetic fields to the magnetic mass. The 4D
SU(2) lattice gauge theory is considered in a deconfinement phase.

The paper is organized as follows. In the next section the
introduction  of chromomagnetic fields on the lattice is
described. In section 3 the results of Monte Carlo~(MC)
simulations are presented. The conclusions and discussion are
given in the last section.

\section{External Chromomagnetic Fields on the Lattice}

 The
Wilson  action for the SU(2) lattice gauge theory is
used~\cite{KS:lit:Gattr}:
\begin{equation}
    \label{KS:eq:WS}
    S=\beta\sum_n\sum_{\mu>\nu}\left[1-\frac12\,\operatorname{Re}\operatorname{Tr} U_{\mu\nu}(n)\right],
\end{equation}
where $\beta=4/g^2$ is inverse coupling, matrix $U_{\mu\nu}$ is
plaquette  in the $\mu\nu$ plane at the point with coordinates
$n=(n_x,n_y,n_z,n_t)$,
\begin{equation}
    \label{KS:eq:plq}
    U_{\mu\nu}(n)=U_\mu(n)U_\nu(n+\hat{\mu})U^\dag_\mu(n+\hat{\nu})U^\dag_\nu(n),
\end{equation}
$\hat{\mu}$ is the unit vector in the $\mu$-th direction, dagger means Hermitian conjugation.

To perform the investigation of the magnetic mass, two
chromomagnetic  fields are introduced. The first one is the field
produced by the monopole-antimonopole string. It is implemented as
described in Refs.~\citen{KS:lit:DGT, KS:lit:SrSu}. The second
field is the constant homogeneous Abelian one, introduced through
the TBC. This method is similar to the one described in
Ref.~\citen{KS:lit:VdF}. It is based on the representation of the
plaquette~(\ref{KS:eq:plq}) through the chromomagnetic field flux
\cite{KS:lit:Gattr}:
\begin{equation}
    \label{KS:eq:UF}
    U_{\mu\nu}(n) = \exp[ia^2F_{\mu\nu}],
\end{equation}
where $F_{\mu\nu}$ is the SU(2) electromagnetic field tensor. If
the   external field is applied along $z$ axis, the plaquette
(\ref{KS:eq:UF}) in $xy$ plane transforms as
\begin{equation}
    \label{KS:eq:HHe}
    U'_{xy}=\exp[ia^2(H_z+H_z^{\mbox{\footnotesize ext}})]=U_{xy}\exp[ia^2 H_z^{\mbox{\footnotesize ext}}]
\end{equation}
up to commutator of $H_z$ and $H_z^{\mbox{\footnotesize ext}}$.
Here $H_z$  is  the stochastic field, which is always present on
the lattice.

\begin{figure}[!htb]
\centering
\includegraphics[bb=46 546 429 698, width=0.45\textwidth]{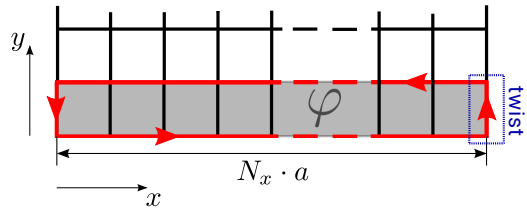}
\caption{Flux of external field $\varphi$ through the stripe of plaquettes}
\label{KS:fig:NxPlqPhi}
\end{figure}

In MC simulations, it is more convenient to attach the external
field to the   link variables instead of plaquettes:
\begin{equation}
    U'_y(n+\hat{x})=U_y(n+\hat{x})\operatorname{e}^{ia^2H_z^{\mbox{\scriptsize ext}}}.
\end{equation}
Here, $H_z^{\mbox{\footnotesize ext}}$ is a constant Abelian
chromomagnetic  field.  In this case the transformation of the
links~(twist) may be performed just on the edge of the lattice.
The choice of the edge slice of the lattice is arbitrary. We make
this transformation at $x=0$,
\begin{equation}
    \label{KS:eq:tbc0}
    U'_y(0,n_y,n_z,n_t)=U_y(0,n_y,n_z,n_t)\operatorname{e}^{i\varphi},
\end{equation}
where
\begin{equation}
    \label{KS:eq:phi}
    \varphi=a^2N_xH_z^{\mbox{\footnotesize ext}},
\end{equation}
represents the  flux of the external field through the
$N_x\times1$ stripe of  plaquettes with coordinates of the left
bottom corner $(0,n_y,n_z,n_t)$, $N_x$ is lattice size in the
$x$-th direction~(see Fig.~\ref{KS:fig:NxPlqPhi}). Since this
transformation is done for all $n_y$, the flux through the whole
$xy$ plane of the lattice is defined.

Accounting for the periodicity  of the lattice, the transformation (\ref{KS:eq:tbc0}) can be rewritten in the form of the TBC:
\begin{equation}
    \left\{
    \begin{array}{l}
        \label{KS:eq:tbc}
        U_y(N_x,n_y,n_z,n_t)=U_y(0,n_y,n_z,n_t)\operatorname{e}^{i\varphi},\\
        U_\mu(N_x,n_y,n_z,n_t)=U_\mu(0,n_y,n_z,n_t),\quad\mu\neq y,\\[1em]
        U_\mu(n_x,N_y,n_z,n_t)=U_\mu(n_x,0,n_z,n_t),\\
        U_\mu(n_x,n_y,N_z,n_t)=U_\mu(n_x,n_y,0,n_t),\\
        U_\mu(n_x,n_y,n_z,N_t)=U_\mu(n_x,n_x,n_z,0).
    \end{array}
    \right.
\end{equation}

In the present paper, the external Abelian field corresponding to the 3-rd Pauli matrix  is introduced. For this case
\begin{equation}
\operatorname{e}^{i\varphi}= \operatorname{e}^{i\varphi_3\sigma_3/2} =\left(\begin{array}{cc}
\operatorname{e}^{i\varphi_3/2} & 0\\
0 & \operatorname{e}^{-i\varphi_3/2}
\end{array}\right),
\end{equation}
where $\sigma_3$ is the 3-rd Pauli matrix, $\varphi_3$ is the
value of  the applied flux. In the following the subscript of the
$\varphi_3$ will be omitted.

\section{Numerical Results}

Let us investigate the influence of the external neutral Abelian
field   on the magnetic mass of the field, produced by the
monopole-antimonopole string. The relevant quantity is the
difference between the mean plaquettes in the presence and in the
absence of the external field:
\begin{equation}
\label{KS:eq:f}
f(N)=|\langle U_{\mbox{\footnotesize field}}\rangle - \langle U_0\rangle|.
\end{equation}
It is considered as a function of lattice size in spatial
direction $N$.  Lattices with symmetric spatial part are used
here. The plaquettes are averaged over the lattice and over some
number of simulation program runs (from 200 up to 1000 for various
lattices).

There are several type behaviors of the quantity $f$ referred in the
literature\cite{KS:lit:DGT}:   $f\sim 1/N^2$, when the magnetic
flux tubes are formed and the flux is conserved; $f\sim1/N^4$,
corresponding to the Coulombic magnetic field, the flux decreases;
$f\sim\operatorname{e}^{-kN^2}$, which means the screening of the
field. In the last case $\sqrt{k}$ is the magnetic mass. From the
investigation of the magnetic mass of the Abelian field \cite{KS:lit:ABDS}, one more possible behavior of $f$ is known: $f\sim1/N$. This corresponds to the increasing of the field flux due to, e.g., spontaneous field generation.

First of all, we are going to reproduce the results of the known
investigation\cite{KS:lit:DGT}. For the field produced by the
monopole-antimonopole string the values $f(N)$ are obtained and
fitted in order to determine, whether the applied field is
screened or not. Then we add the external Abelian field flux and
look how it changes the behavior of the quantity $f(N)$.

The MC simulations are performed on lattices $4\times N^3$ at
three  inverse couplings $\beta=2.835,\ 3.020,\ 3.091$, which
correspond to the temperatures about 1.2, 1.9, 2.3~GeV. These
temperatures are estimated using the relation\cite{KS:lit:CB}
\begin{equation}
    T = T_c\,\frac{a(\beta_c)}{a(\beta)},
\end{equation}
where critical $\beta$ for the $4\times N^3$ lattices
$\beta_c=2.301$  is found at $4\times24^3$ lattice using  the
Polyakov loop susceptibility, $T_c=0.313$~GeV is taken from
Ref.~\citen{KS:lit:BCLP}. Note that $\beta=3.020$ corresponds to
the temperature used in Ref.~\citen{KS:lit:DGT}~(in that case
$\beta_c=1.8735$ is taken from Ref.~\citen{KS:lit:Vel}).

The simulations are carried out with modification of the QCDGPU
program  \cite{KS:lit:QCDGPU} at GPUs of HybriLIT
\cite{KS:lit:hlit} cluster, HPCVillage \cite{KS:lit:hpcv} and HGPU
Group \cite{KS:lit:hgpu} machines. In all the simulations the
multihit heat-bath algorithm is used to  update links, the number
of hits is 10. Pseudorandom numbers are generated with RANLUX3
generator. Each program run includes~2000 thermalization sweeps
and~1000 sweeps with measurements separated by~9 sweeps without
measurements for decorrelation. The cold start is used.
Calculations are performed with double precision.

The quantities $\langle U_{\mbox{\footnotesize field}}\rangle$ and
$\langle U_0\rangle$  are calculated for each lattice size
separately and combined into the quantity $f(N)$. To discern the
functional dependence describing obtained data, they are
linearized by logarithmization and fitted through minimization of
$\chi^2$ function,
\begin{equation}
\label{KS:eq:chi2}
\chi^2(a)=\sum_{i=1}^K\frac{[y_i-\log f(N_i;a)]^2}{\sigma_i^2},
\end{equation}
where $y_i=\log f_i$, $f_i$ are the data for the quantity
(\ref{KS:eq:f})  obtained from simulations, $f(N;a)$ is a trial
function, $K$ is the number of points used in fit, $a$ is a set of
fitting parameters. The statistical errors~$\sigma_i$ are
estimated as for indirect measurement. Combinations of inverse
power and exponential functions are tried in fit. To select which
functions may describe data and which ones should be discarded,
the $\chi^2$ criterion is used. The null hypothesis is that the
trial function does describe the data, the alternate hypothesis is
that it does not describe them. For each fit function the minimal
value $\chi^2_{min}$ is calculated by Eq.\,\ref{KS:eq:chi2} at
estimated parameters $a$. This quantity is independent of the input parameters and follows the chi-squared distribution with~$\nu$ degrees of freedom,
\begin{equation}
\nu = K - L,
\end{equation}
where $L$ is the number of estimated parameters. Assuming that
large values  of $\chi^2_{min}$ are unlikely, it is compared with
critical value of the chi-squared distribution
$\chi^2_{\nu,0.05}$. If $\chi^2_{min} > \chi^2_{\nu,0.05}$, the
null hypothesis is rejected. Here 0.05 is the probability of
rejection of the true null hypothesis~(type I error).

It can be shown that the difference
\begin{equation}
\label{KS:eq:dchi2}
\Delta\chi^2=\chi^2(a) - \chi^2_{min}
\end{equation}
is independent of the estimating and the input parameters and has
the chi-squared distribution with~$L$ degrees of freedom. Thus,
the quantity (\ref{KS:eq:dchi2}) is compared with the critical
value $\chi^2_{L,0.05}$, and the confidence intervals (CIs) for
the parameters $a$ at 95\%~confidence level~(CL) can be found from
inequality
\begin{equation}
\chi^2(a) \leq \chi^2_{min} + \chi^2_{L,0.05}.
\end{equation}

\begin{figure}[!htb]
    \centering
    \includegraphics[bb=18 180 594 612, width=0.6\textwidth]{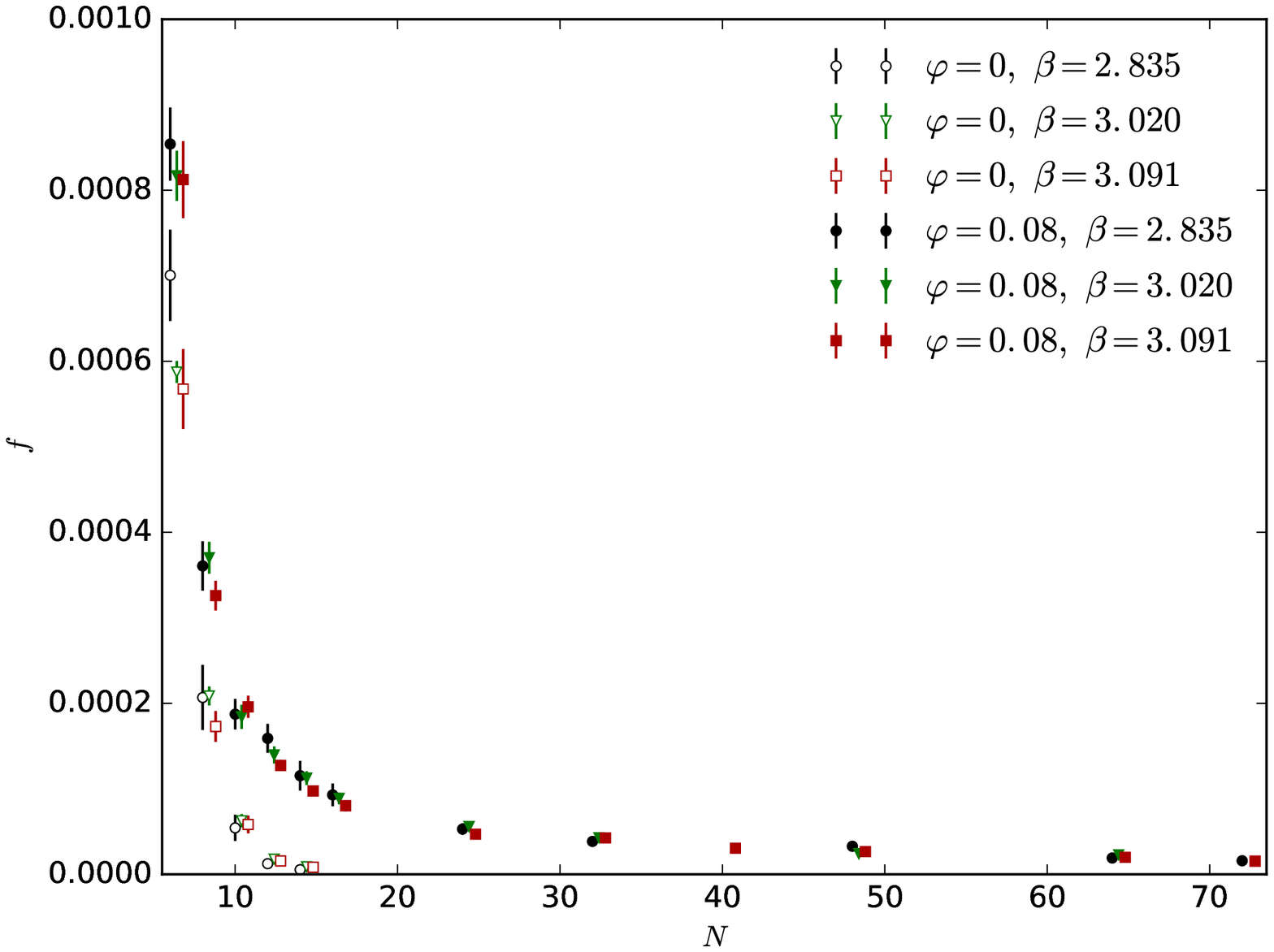}\\
    \includegraphics[bb=18 180 594 612, width=0.6\textwidth]{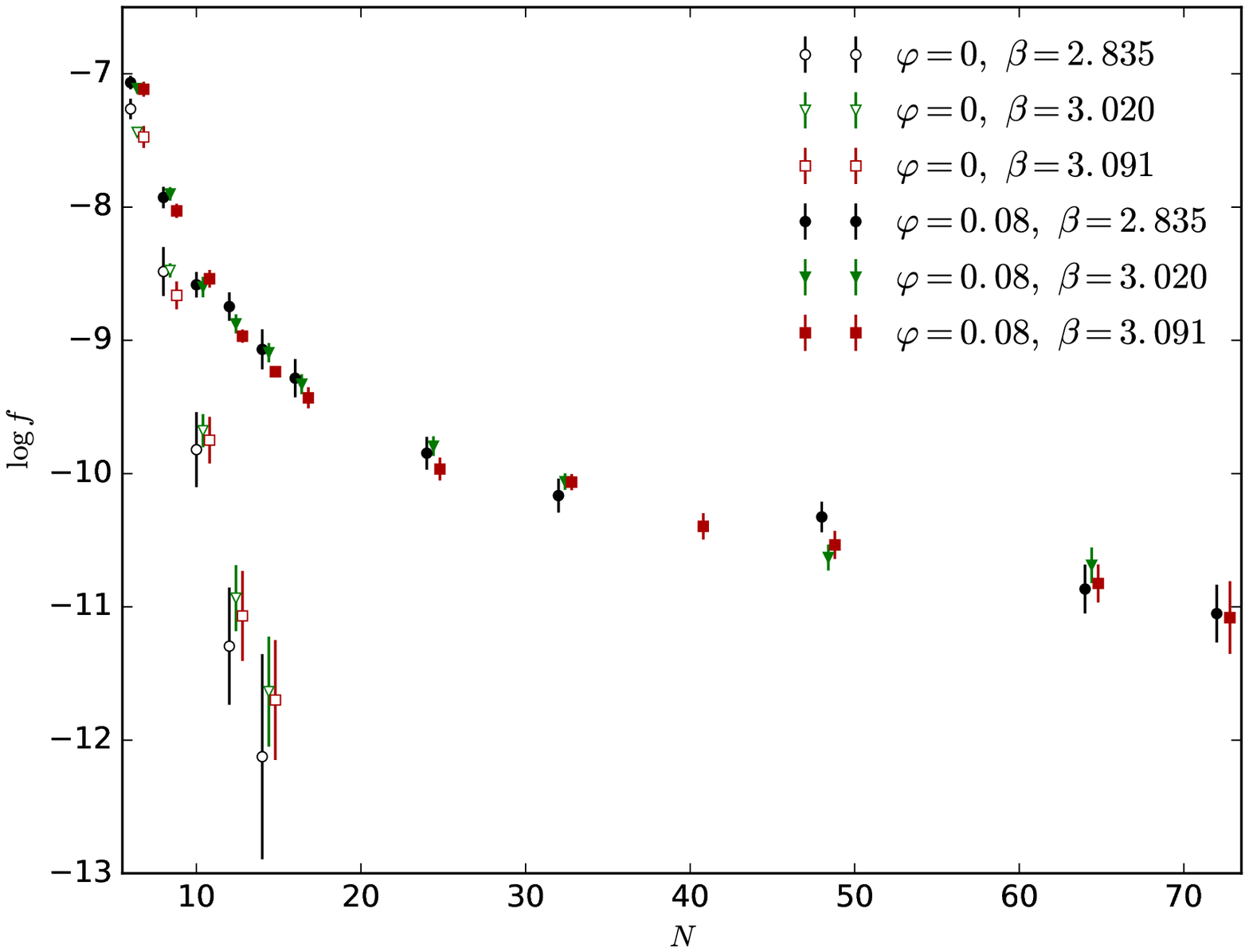}
    \caption{%
    (top) Contributions of the external field to the average plaquette in the cases of absence~(open markers) and presence~(filled markers) of the additional Abelian field flux; (bottom) logarithms of these contributions. The bins represent standard deviations for the measured quantities. For $\beta=3.020$ and $\beta=3.091$ data are shifted to the right for better visibility.
    \label{KS:fig:data_f}}
\end{figure}

The data are presented in Fig. \ref{KS:fig:data_f}, the results of
fitting are shown in Tables~\ref{KS:tab:fit_res0-1} --
\ref{KS:tab:fit_res0-3} for three considered temperatures. In all
these tables the first columns contain trial fit functions, the
second columns show the minimal $\chi^2$ values, the fourth
columns contain threshold values $\chi^2_{\nu,0.05}$ for $\nu$
degrees of freedom listed in the third columns, then the
conclusions from hypothesis testing are shown, and the last
columns represent 95\%~CIs for parameters responsible for
screening got from fit (dimensionless).

\begin{table}[pthb]
\tbl{Fit results at $\varphi = 0$, $\beta=2.835$}
{%
\begin{tabular}{@{}lccccc@{}}
\toprule
\multicolumn{1}{c}{\scriptsize Function} & \scriptsize$\hspace{2em}\chi^2_{min}$\hspace{2em} & $\nu$ & $\chi^2_{\nu,0.05}$ & \scriptsize Conclusion & \scriptsize Estimate of $k$ with $2\sigma$\,CI\\
\colrule
$A/N^2$ & 80.4 & 4 & 9.49 & rejected & --\rule[-0.6em]{0pt}{1.2em}\\
$\left(A/N^2\right)\operatorname{e}^{-kN}$ & 0.99 & 3 & 7.81 & accepted & $\left(3.95\pm1.09\right)\times10^{-1}$\rule{0pt}{1.2em}\rule[-0.6em]{0pt}{1.2em}\\
$\left(A/N^2\right)\operatorname{e}^{-kN^2}$ & 0.63 & 3 & 7.81 & accepted & $\left(2.30\pm0.63\right)\times10^{-2}$\rule{0pt}{1.4em}\rule[-0.7em]{0pt}{1.4em}\\
\hline
$A/N^4$ & 14.0 & 4 & 9.49 & rejected & --\rule{0pt}{1.2em}\rule[-0.6em]{0pt}{1.2em}\\
$\left(A/N^4\right)\operatorname{e}^{-kN}$ & 2.32 & 3 & 7.81 & accepted & $(1.51\pm1.09)\times10^{-1}$\rule{0pt}{1.2em}\rule[-0.6em]{0pt}{1.2em}\\
$\left(A/N^4\right)\operatorname{e}^{-kN^2}$ & 1.36 & 3 & 7.81 & accepted & $(9.13\pm6.29)\times10^{-3}$\rule{0pt}{1.4em}\rule[-0.7em]{0pt}{1.4em}\\
\hline
$A\operatorname{e}^{-kN}$ & 0.40 & 3 & 7.81 & accepted & $(6.39\pm1.09)\times10^{-1}$\rule{0pt}{1.2em}\rule[-0.6em]{0pt}{1.2em}\\
$A\operatorname{e}^{-kN^2}$ & 3.18 & 3 & 7.81 & accepted & $(3.68\pm0.63)\times10^{-2}$\rule{0pt}{1.4em}\rule[-0.7em]{0pt}{1.4em}\\
\hline
$A/N$ & 137 & 4 & 9.49 & rejected & --\rule{0pt}{1.2em}\rule[-0.6em]{0pt}{1.2em}\\
$(A/N)\operatorname{e}^{-kN}$ & 0.60 & 3 & 7.81 & accepted & $(5.17\pm1.09)\times10^{-1}$\rule{0pt}{1.2em}\rule[-0.6em]{0pt}{1.2em}\\
$(A/N)\operatorname{e}^{-kN^2}$ & 1.49 & 3 & 7.81 & accepted & $(2.99\pm0.63)\times10^{-2}$\rule{0pt}{1.4em}\\
\botrule
\end{tabular}
\label{KS:tab:fit_res0-1}}
\end{table}

\begin{table}[pthb]
\tbl{Fit results at $\varphi = 0$, $\beta=3.020$}
{%
\begin{tabular}{@{}lccccc@{}}
\toprule
\multicolumn{1}{c}{\scriptsize Function} & \scriptsize$\hspace{2em}\chi^2_{min}$\hspace{2em} & $\nu$ & $\chi^2_{\nu,0.05}$ & \scriptsize Conclusion & \scriptsize Estimate of $k$ with $2\sigma$\,CI\\
\colrule
$A/N^2$ & 247 & 4 & 9.49 & rejected & --\rule[-0.6em]{0pt}{1.2em}\\
$\left(A/N^2\right)\operatorname{e}^{-kN}$ & 7.29 & 3 & 7.81 & accepted & $\left(2.86\pm0.45\right)\times10^{-1}$\rule{0pt}{1.2em}\rule[-0.6em]{0pt}{1.2em}\\
$\left(A/N^2\right)\operatorname{e}^{-kN^2}$ & 2.16 & 3 & 7.81 & accepted & $\left(1.78\pm0.28\right)\times10^{-2}$\rule{0pt}{1.4em}\rule[-0.7em]{0pt}{1.4em}\\
\hline
$A/N^4$ & 19.5 & 4 & 9.49 & rejected & --\rule{0pt}{1.2em}\rule[-0.6em]{0pt}{1.2em}\\
$\left(A/N^4\right)\operatorname{e}^{-kN}$ & 17.2 & 3 & 7.81 & rejected & $(2.80\pm4.52)\times10^{-2}$\rule{0pt}{1.2em}\rule[-0.6em]{0pt}{1.2em}\\
$\left(A/N^4\right)\operatorname{e}^{-kN^2}$ & 15.5 & 3 & 7.81 & rejected & $(2.28\pm2.78)\times10^{-3}$\rule{0pt}{1.4em}\rule[-0.7em]{0pt}{1.4em}\\
\hline
$A\operatorname{e}^{-kN}$ & 2.19 & 3 & 7.81 & accepted & $(5.44\pm0.45)\times10^{-1}$\rule{0pt}{1.2em}\rule[-0.6em]{0pt}{1.2em}\\
$A\operatorname{e}^{-kN^2}$ & 11.8 & 3 & 7.81 & rejected & $(3.33\pm0.28)\times10^{-2}$\rule{0pt}{1.4em}\rule[-0.7em]{0pt}{1.4em}\\
\hline
$A/N$ & 509 & 4 & 9.49 & rejected & --\rule{0pt}{1.2em}\rule[-0.6em]{0pt}{1.2em}\\
$(A/N)\operatorname{e}^{-kN}$ & 4.14 & 3 & 7.81 & accepted & $(4.15\pm0.45)\times10^{-1}$\rule{0pt}{1.2em}\rule[-0.6em]{0pt}{1.2em}\\
$(A/N)\operatorname{e}^{-kN^2}$ & 4.09 & 3 & 7.81 & accepted & $(2.55\pm0.28)\times10^{-2}$\rule{0pt}{1.4em}\\
\botrule
\end{tabular}
\label{KS:tab:fit_res0-2}}
\end{table}

\begin{table}[pthb]
\tbl{Fit results at $\varphi = 0$, $\beta=3.091$}
{%
\begin{tabular}{@{}lccccc@{}}
\toprule
\multicolumn{1}{c}{\scriptsize Function} & \scriptsize$\hspace{2em}\chi^2_{min}$\hspace{2em} & $\nu$ & $\chi^2_{\nu,0.05}$ & \scriptsize Conclusion & \scriptsize Estimate of $k$ with $2\sigma$\,CI\\
\colrule
$A/N^2$ & 102 & 4 & 9.49 & rejected & --\rule[-0.6em]{0pt}{1.2em}\\
$\left(A/N^2\right)\operatorname{e}^{-kN}$ & 0.77 & 3 & 7.81 & accepted & $\left(3.27\pm0.80\right)\times10^{-1}$\rule{0pt}{1.2em}\rule[-0.6em]{0pt}{1.2em}\\
$\left(A/N^2\right)\operatorname{e}^{-kN^2}$ & 1.89 & 3 & 7.81 & accepted & $\left(1.85\pm0.45\right)\times10^{-2}$\rule{0pt}{1.4em}\rule[-0.7em]{0pt}{1.4em}\\
\hline
$A/N^4$ & 9.53 & 4 & 9.49 & rejected & --\rule{0pt}{1.2em}\rule[-0.6em]{0pt}{1.2em}\\
$\left(A/N^4\right)\operatorname{e}^{-kN}$ & 2.45 & 3 & 7.81 & accepted & $(8.65\pm7.96)\times10^{-2}$\rule{0pt}{1.2em}\rule[-0.6em]{0pt}{1.2em}\\
$\left(A/N^4\right)\operatorname{e}^{-kN^2}$ & 1.58 & 3 & 7.81 & accepted & $(5.22\pm4.53)\times10^{-3}$\rule{0pt}{1.4em}\rule[-0.7em]{0pt}{1.4em}\\
\hline
$A\operatorname{e}^{-kN}$ & 0.98 & 3 & 7.81 & accepted & $(5.68\pm0.80)\times10^{-1}$\rule{0pt}{1.2em}\rule[-0.6em]{0pt}{1.2em}\\
$A\operatorname{e}^{-kN^2}$ & 10.3 & 3 & 7.81 & rejected & $(3.18\pm0.45)\times10^{-2}$\rule{0pt}{1.4em}\rule[-0.7em]{0pt}{1.4em}\\
\hline
$A/N$ & 190 & 4 & 9.49 & rejected & --\rule{0pt}{1.2em}\rule[-0.6em]{0pt}{1.2em}\\
$(A/N)\operatorname{e}^{-kN}$ & 0.64 & 3 & 7.81 & accepted & $(4.48\pm0.80)\times10^{-1}$\rule{0pt}{1.2em}\rule[-0.6em]{0pt}{1.2em}\\
$(A/N)\operatorname{e}^{-kN^2}$ & 5.10 & 3 & 7.81 & accepted & $(2.52\pm0.45)\times10^{-2}$\rule{0pt}{1.4em}\\
\botrule
\end{tabular}
\label{KS:tab:fit_res0-3}}
\end{table}

As it is seen,  our results qualitatively reproduce the result of
Ref.~\citen{KS:lit:DGT}. Fit  functions without screening factor
give large $\chi^2_{min}$, so they are excluded. Function
$A\operatorname{e}^{-kN^2}$ considered in Ref.~\citen{KS:lit:DGT}
is more preferable, but it also gives
$\chi^2_{min}>\chi^2_{\nu,0.05}$ for some temperatures. Thus, this
behavior is also excluded at 95\%~CL for these temperatures. Usage
of $N$-dependent prefactor improves the goodness of fit. It worth to mention, that all accepted trial functions are indistinguishable at the present data, despite they describe different physics.

Then the external neutral Abelian field flux is added. It is directed  along the monopole-antimonopole string. The value of $\varphi=0.08$ is chosen such that it is small enough to be used in TBC and large enough to influence the average plaquette~(checked at $2\sigma$ significance level).

The data in presence of both the monopole-antimonopole and
additional  Abelian field flux are shown in~Fig.\,\ref{KS:fig:data_f}. Points with filled markers correspond to them. It is seen that addition of the neutral
Abelian field flux significantly changes the behavior of the
measured quantity~(\ref{KS:eq:f})~-- it decreases much slower than
in the   absence of the Abelian field. To describe the behavior of
$f$, the logarithms of the data in the presence of the Abelian
field are also fitted by means of $\chi^2$ method. Fit results are
presented in Tables~\ref{KS:tab:fit_resphi-1} -- \ref{KS:tab:fit_resphi-3} for each temperature. The first three fit functions are the generalization of functions used in Tables~\ref{KS:tab:fit_res0-1} -- \ref{KS:tab:fit_res0-3}. The
fourth function is motivated by observation that the
behavior of the data resembles the typical dependence of static
quark potential on distance\cite{KS:lit:NS,KS:lit:Gattr}, up to
the signs of the terms~(see Fig.\,\ref{KS:fig:data_f} (bottom)).

\begin{table}[pthb]
\tbl{Fit results at $\varphi = 0.08$, $\beta=2.835$}
{%
\begin{tabular}{@{}lccccc@{}}
\toprule
\multicolumn{1}{c}{\scriptsize Function} & \scriptsize$\hspace{0.25em}\chi^2_{min}$\hspace{0.25em} & \scriptsize$\hspace{0.01em}\nu$\hspace{0.01em} & \scriptsize$\hspace{0.25em}\chi^2_{\nu, 0.05}$\hspace{0.25em} & \scriptsize Conclusion & \scriptsize Estimate of parameters with $2\sigma$\,CI\\
\colrule
$A/N^b$ & 91.2 & 9 & 16.9 & rejected & $b=1.65\pm0.10$\rule{0pt}{0.8em}\rule[-0.7em]{0pt}{1.4em}\\
\hline
$(A/N^b)\operatorname{e}^{-kN}$ & 30.0 & 8 & 15.5 & rejected & $b=2.59\pm0.35$\rule{0pt}{1.4em}\rule[-0.7em]{0pt}{1.4em}\\
&&&&& $k=(-4.59\pm1.63)\times10^{-2}$\rule{0pt}{1.4em}\rule[-0.7em]{0pt}{1.4em}\\
\hline
$(A/N^b)\operatorname{e}^{-kN^2}$ & 47.2 & 8 & 15.5 & rejected & $b=2.07\pm0.20$\rule{0pt}{1.4em}\rule[-0.7em]{0pt}{1.4em}\\
&&&&& $k=(-3.45\pm1.43)\times10^{-4}$\rule{0pt}{1.4em}\rule[-0.7em]{0pt}{1.4em}\\
\hline
$A\operatorname{e}^{B/N}\operatorname{e}^{-kN}$ & 5.33 & 8 & 15.5 & accepted & $B=20.3\pm2.64$\rule{0pt}{1.4em}\rule[-0.7em]{0pt}{1.4em}\\
&&\phantom{10}&&& $k=(1.09\pm0.91)\times10^{-2}$\rule{0pt}{1.4em}\rule[-0.7em]{0pt}{1.4em}\\
\botrule
\end{tabular}
\label{KS:tab:fit_resphi-1}}
\end{table}

\begin{table}[pthb]
\tbl{Fit results at $\varphi = 0.08$, $\beta=3.020$}
{%
\begin{tabular}{@{}lccccc@{}}
\toprule
\multicolumn{1}{c}{\scriptsize Function} & \scriptsize$\hspace{0.25em}\chi^2_{min}$\hspace{0.25em} & \scriptsize$\hspace{0.01em}\nu$\hspace{0.01em} & \scriptsize$\hspace{0.25em}\chi^2_{\nu, 0.05}$\hspace{0.25em} & \scriptsize Conclusion & \scriptsize Estimate of parameters with $2\sigma$\,CI\\
\colrule
$A/N^b$ & 170 & 8 & 15.5 & rejected & $b=1.71\pm0.07$\rule{0pt}{0.8em}\rule[-0.7em]{0pt}{1.4em}\\
\hline
$(A/N^b)\operatorname{e}^{-kN}$ & 44.9 & 7 & 14.1 & rejected & $b=2.65\pm0.24$\rule{0pt}{1.4em}\rule[-0.7em]{0pt}{1.4em}\\
&&&&& $k=(-5.22\pm1.29)\times10^{-2}$\rule{0pt}{1.4em}\rule[-0.7em]{0pt}{1.4em}\\
\hline
$(A/N^b)\operatorname{e}^{-kN^2}$ & 73.7 & 7 & 14.1 & rejected & $b=2.11\pm0.14$\rule{0pt}{1.4em}\rule[-0.7em]{0pt}{1.4em}\\
&&&&& $k=(-4.44\pm1.25)\times10^{-4}$\rule{0pt}{1.4em}\rule[-0.7em]{0pt}{1.4em}\\
\hline
$A\operatorname{e}^{B/N}\operatorname{e}^{-kN}$ & 7.14 & 7 & 14.1 & accepted & $B=20.1\pm1.84$\rule{0pt}{1.4em}\rule[-0.7em]{0pt}{1.4em}\\
&&\phantom{10}&&& $k=(1.08\pm0.75)\times10^{-2}$\rule{0pt}{1.4em}\rule[-0.7em]{0pt}{1.4em}\\
\botrule
\end{tabular}
\label{KS:tab:fit_resphi-2}}
\end{table}

\begin{table}[pthb]
\tbl{Fit results at $\varphi = 0.08$, $\beta=3.091$}
{%
\begin{tabular}{@{}lccccc@{}}
\toprule
\multicolumn{1}{c}{\scriptsize Function} & \scriptsize$\hspace{0.25em}\chi^2_{min}$\hspace{0.25em} & \scriptsize$\hspace{0.01em}\nu$\hspace{0.01em} & \scriptsize$\hspace{0.25em}\chi^2_{\nu, 0.05}$\hspace{0.25em} & \scriptsize Conclusion & \scriptsize Estimate of parameters with $2\sigma$\,CI\\
\colrule
$A/N^b$ & 223 & 10 & 18.3 & rejected & $b=1.56\pm0.08$\rule{0pt}{0.8em}\rule[-0.7em]{0pt}{1.4em}\\
\hline
$(A/N^b)\operatorname{e}^{-kN}$ & 69.0 & 9 & 16.9 & rejected & $b=2.67\pm0.26$\rule{0pt}{1.4em}\rule[-0.7em]{0pt}{1.4em}\\
&&&&& $k=(-5.55\pm1.23)\times10^{-2}$\rule{0pt}{1.4em}\rule[-0.7em]{0pt}{1.4em}\\
\hline
$(A/N^b)\operatorname{e}^{-kN^2}$ & 118 & 9 & 16.9 & rejected & $b=2.02\pm0.15$\rule{0pt}{1.4em}\rule[-0.7em]{0pt}{1.4em}\\
&&&&& $k=(-4.30\pm1.16)\times10^{-4}$\rule{0pt}{1.4em}\rule[-0.7em]{0pt}{1.4em}\\
\hline
$A\operatorname{e}^{B/N}\operatorname{e}^{-kN}$ & 7.00 & 9 & 16.9 & accepted & $B=21.3\pm2.01$\rule{0pt}{1.4em}\rule[-0.7em]{0pt}{1.4em}\\
&&&&& $k=(6.90\pm6.76)\times10^{-3}$\rule{0pt}{1.4em}\rule[-0.7em]{0pt}{1.4em}\\
\botrule
\end{tabular}
\label{KS:tab:fit_resphi-3}}
\end{table}

\begin{figure}[!htb]
    \centering
    \includegraphics[bb=18 180 594 612, width=0.48\textwidth]{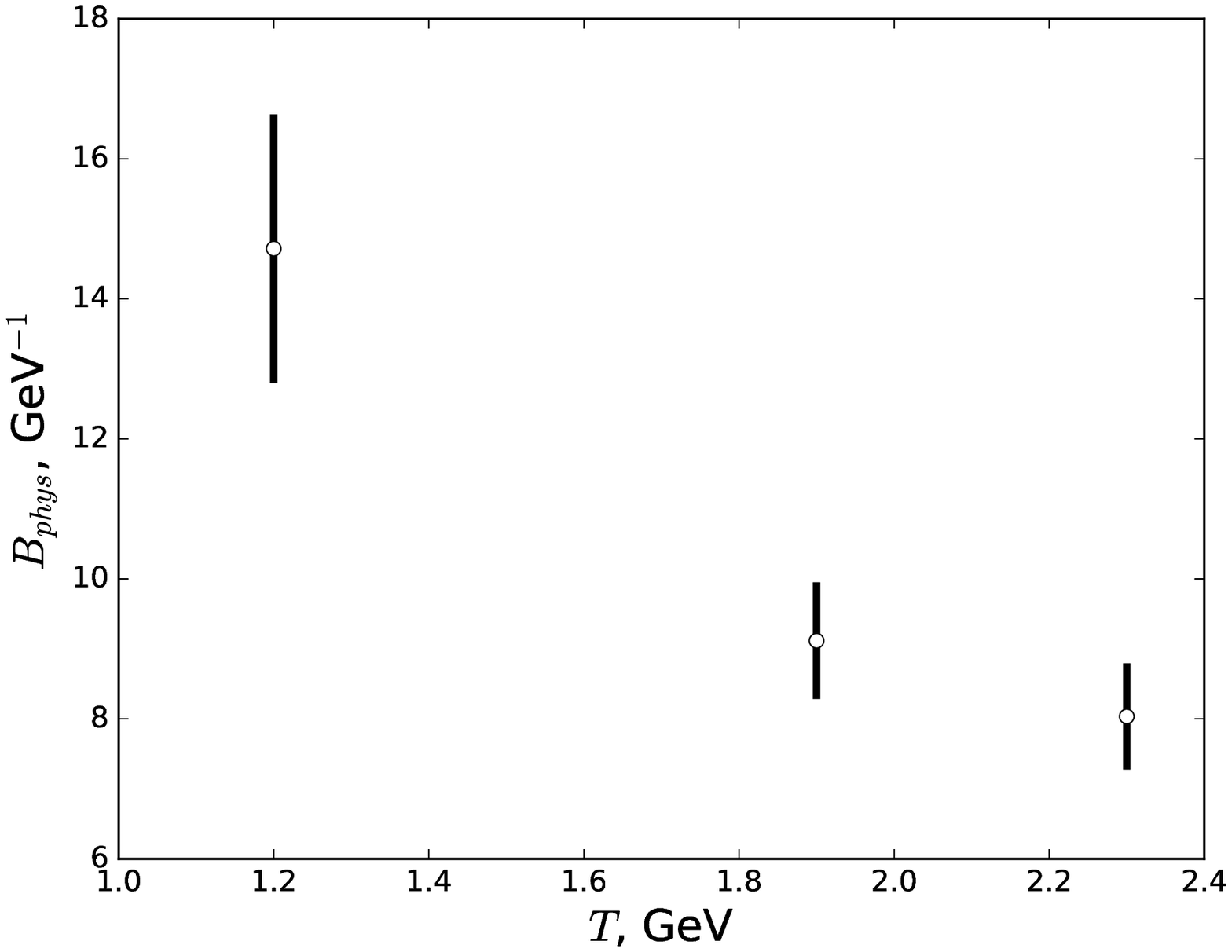}~%
    \includegraphics[bb=18 180 594 612, width=0.48\textwidth]{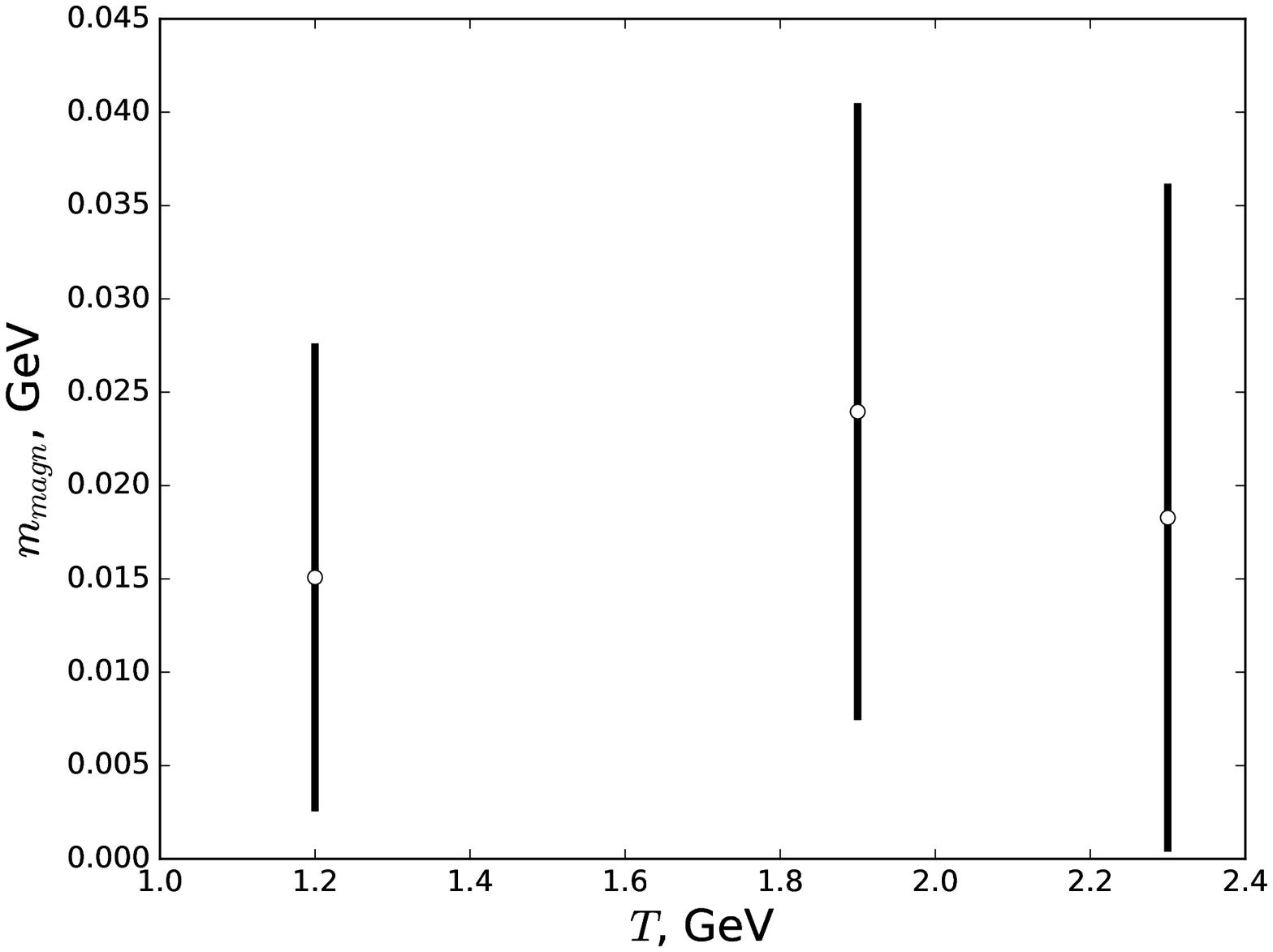}
    \caption{Quantities $B_{phys}$ and $m_{magn}$ as functions of temperature. Bars correspond to~95\%~CIs.
    \label{KS:fig:BkT}}
\end{figure}

As it is seen from the tables, for all three temperatures the
first three trial  functions are rejected at 95\% CL. The last
behavior is accepted at this CL. In the correspondent function
\begin{equation}\label{KS:eq:bf}
f=A\operatorname{e}^{B/N}\operatorname{e}^{-kN}
\end{equation}
the parameter $k$ serves as magnetic mass, while $B$ is some ``enhancing''  parameter. Passing to the physical units, one can see that the parameter
\begin{equation}
    B_{phys} = a\cdot B
\end{equation}
decreases with temperature, while the magnetic mass
\begin{equation}
    m_{magn} = k / a
\end{equation}
is constant within CIs at the considered range of temperatures
(see Fig.\,\ref{KS:fig:BkT}); $a$ is lattice spacing. Averaging over three used temperatures one obtains the magnetic mass $(1.83\pm0.87)\times10^{-2}$\,GeV at 95\%~CL.

To compare the obtained magnetic mass values with the ones corresponding  to the absence of the additional Abelian flux, the data at $\varphi = 0$ are fitted with the trial function (\ref{KS:eq:bf}). The results of this fit are shown in
Table~\ref{KS:tab:fit_res0-4}. The first line represents minimal
$\chi^2$ values. Comparison of them with $\chi^2_{\nu; 0.05}=5.99$
shows that the function~(\ref{KS:eq:bf}) describes data correctly at
95\%~CL for all three temperatures. The last two rows contain the
estimates and CIs for the dimensionless parameters~$B$ and~$k$. The averaged
over considered temperatures  magnetic mass is
$1.26\pm0.41$\,GeV.

\begin{table}[pthb]
\tbl{Fit results at $\varphi = 0$ with
$f=A\operatorname{e}^{B/N}\operatorname{e}^{-kN}$.  For all
temperatures $\nu=2$, $\chi^2_{\nu;0.05}=5.99$}
{%
\begin{tabular}{@{}lccc@{}}
\toprule
& $\beta=2.835$ & $\beta=3.020$ & $\beta=3.091$\\
\hline
$\chi^2_{min}$ & 0.36 & 1.34 & 0.64\\
$\hat{B}\pm 2\sigma$ CI & $-2.81\pm37.6$ & $-4.95\pm14.2$ & $5.04\pm23.9$\\
$\hat{k}\pm 2\sigma$ CI & $(6.83\pm5.76)\times10^{-1}$ & $(6.29\pm2.46)\times10^{-1}$ & $(4.92\pm3.66)\times10^{-1}$\\
\botrule
\end{tabular}
\label{KS:tab:fit_res0-4}}
\end{table}

Hence it is seen that the magnetic mass of the chromomagnetic field produced by monopole-antimonopole string in the presence of the external Abelian field flux is two orders of magnitude less than the magnetic mass of solely
monopole-antimonopole string field. This means that the magnetic mass of the neutral Abelian field is much smaller than the magnetic mass of the color charged components.

Let us discuss this in more details. To simplify situation we
propose that the  first field flux produced by the string through
one plaquette    is described by  the function $e^{- a N}$ and for
the second one generated  by TBC it  is $e^{- b N}$, $a$ and $b$ are
corresponding magnetic masses. The magnetic mass of the total flux
is defined from the function  $e^{- c  N}$. Hence, independently
of the  chosen prefactors, the relation holds
\be \label{c} e^{- a  N} + e^{- b N} = 2 e^{- c  N} . \ee It
follows from the flux conservation. Assuming the masses are small,
it is easily to get the relation  $ c = \sqrt{a b}$.  To have $c$
two order less than $a$ the mass  $b$ has to be of the order $b
\sim a^5$. That is
 the magnetic mass of the Abelian magnetic field must  be negligibly small and
dominant in the total flux. This conclusion is in agreement with
the results of  Refs. \citen{KS:lit:BS2007,
KS:lit:BS2008}, \citen{KS:lit:SS2005}, \citen{KS:lit:ABDS} obtained in analytic and MC calculations, that this mass is zero. Thus, there is no static
screening effect for such type fields.
This fact  is similar to the one for  usual static magnetic field in QED plasma.

\section{Conclusions and Discussion}

To investigate the SU(2) structure of the magnetic mass, the
influence of the  external Abelian neutral chromomagnetic field on
the field of monopole-antimonopole string was investigated by the
methods of lattice gluodynamics.

First we convinced that in the absence of the external Abelian field  the
results of  Ref.~\citen{KS:lit:DGT} are reproduced. We have
obtained that the field of the  monopole-antimonopole string is
screened. When the external neutral Abelian field flux is added,
the character of screening is changed drastically.  It was found that
it is not  described by just a  one decreasing exponent anymore.
The extra factor corresponding to ``enhancement'' behavior
appears. This factor decreases with temperature increases, whereas
the magnetic mass turns out to be two
order less than at zero Abelian flux and constant for all the
range of considered temperatures.

Appearance of the ``enhancing'' factor demonstrates (independently of
Ref.~\citen{KS:lit:ABDS}) that magnetic mass of the Abelian field
is zero with high accuracy possible in numeric calculations.
Moreover, the magnetic flux passing through a plaquette is grater
compared to the zero field case. This also means that Abelian
field  occupies whole the volume of the lattice.  On the
contrary, the non-Abelian field is screened, and it may be
observed just up to distances of the order of the inverse magnetic
mass in Eq.~(\ref{KS:eq:mcolan}). Thus, as a conclusion,   the Abelian magnetic
fields only  have to be present in the volume of QGP. To investigate
in detail  the analytic field  and temperature dependence of
$m_{magn}(H, T)$ one has to consider a model with more then one
elements in the center of the group. This is a subject for further
investigation.

The absence of the static screening for Abelian magnetic fields
resembles the case of magnetic fields in QED. In gluodynamics, it
has also been detected already in analytic \cite{KS:lit:BS2007,
KS:lit:BS2008, KS:lit:SS2005} and lattice \cite{KS:lit:ABDS}
calculations.  This is in contrast to the Debye mass. In the field
presence  the mass  is (in high temperature approximation)\cite{KS:lit:BS2007}:
\be \label{mD} m^2_D(H, T) = \frac{2}{3} T^2 \left[1 - 0.8859
\left(\frac{\sqrt{g H}}{2 T} \right) + O ((g H)^2/T^4)\right]. \ee
Its value is decreased compared to the zero field case. Thus,
 chromoelectric static fields become more long
range ones.

However, such a situation does not exclude  a dynamical screening
and gap for gluon modes. It can happen for nonzero momentum $k_4
\not = 0$. In QGP  under influence of strong magnetic fields, it
was investigated recently in Ref.~\citen{KS:lit:HS} on the base of
Schwinger-Dyson's equation for gluon fields.

The Abelian field was introduced on the lattice through the TBC.
Despite  this field is not gauge invariant, it is the solution to
the Yang-Mills  field equations without  sources. Just such type
solutions can be realized spontaneously in nature. How to deal
with the non-gauge invariant solutions  to obtain a gauge
invariant vacuum has been discussed by Feynman \cite{KS:lit:Feyn}.
He noted  that to find a gauge invariant vacuum one should  select
a contour in space-time which saplies a gauge invariant integral
for a chosen field potential. Such procedure leads to the vacuum
domain structure. The form of the lattice can be obtained from the
requirements of gauge invariance for phase integral over lattice
contour and lowering the free energy of this configuration. Thus,
it is expected the lattice vacuum structure of QGP. In fact, it is
not difficult to investigate such a structure on the lattice and
this is also the problem  for the future.

\section*{Acknowledgments}

The authors are grateful to  Alexey~V.~Gulov for useful
discussions  and careful reading the manuscript. One of us
would like to thank the HybriLIT of JINR, HPC Village  project and
HGPU group for the computational resources provided.

\bibliographystyle{ws-ijmpa}
\bibliography{KolomoyetsSkalozub}
\end{document}